\documentclass[twoside,10pt,a4paper]{newFNLstyle}
\usepackage{graphics}
\usepackage{cite}

\begin{document}

\volnumpagesyear{0}{0}{000--000}{2001}
\dates{Received 25 January 2005}{Revised 24 February 2005}{Accepted 1 March 2005}

\title{On Fermion Entanglement}

\authorsone{T. A. Kaplan}
\affiliationone{Department of Physics and Astronomy and Institute for Quantum Sciences \\ Michigan State
University, East Lansing, Michigan 48824}


\maketitle




\begin{abstract}
  Under the historical definition of entanglement, namely that
encountered in Einstein--Podolski--Rosen (EPR)--type experiments, it is shown that a particular Slater
determinant is entangled. Thus a definition which holds that any Slater determinant is unentangled, found in the
literature, is inconsistent with the historical definition. A generalization of the historical definition that
embodies its meaning as quantum correlation of observables, in the spirit of the work of many others, but is
much simpler and more physically transparent, is presented.
\end{abstract}
\vspace{.25in} The concept of entanglement is important to the conceptual structure of quantum computation and
information transfer, as seen by the many recent papers dealing with it in this context, e.g.,~\cite{bennett,
zanardi01, zanardi02,zanardi02', zanardi04, shi, schliemann, eckert, barnum0, barnum,  ortiz}. The term
entanglement was coined by Schr\"{o}dinger~\cite{schrodinger} regarding the essentially non-classical behavior
discussed in the famous Gedanken experiment of Einstein, Podolski and Rosen (EPR).~\cite{einstein} The original
concept, as simplified by Bohm~\cite{bohm}, is as follows. Consider two spin-1/2 particles, e.g. two atoms in a
molecule or (I add) the two electrons in a He atom, such that the system is initially in its ground state with
total spin $S=0$, i.e. the singlet. Then the force that binds them together is removed, the particles moving
apart so that their interaction becomes negligible. Despite the latter, a measurement of the spin of one of the
particles (undeterminable before the measurement!) will immediately predict what a spin measurement on the other
one will yield. These correlations differ markedly from those expected in classical mechanics, where each spin
can be known at all times. The essential difference is expressed technically by the fact that
quantum-mechanically the two spins can be correlated in the usual sense, $<s_1^\zeta
s_2^\zeta>\ne<s_1^\zeta><s_2^\zeta>$ (as in the singlet state). Whereas \textit{classically} $<s_1^\zeta
s_2^\zeta>=<s_1^\zeta><s_2^\zeta>,$ the variables, accordingly, being said to be uncorrelated. Here $<\cdot>$
denotes a pure state average. This is at one instant of time, so, each spin classically having a definite value,
averages of functions of the spins are simply those functions. I.e., at a given time there are no fluctuations
in the variables, whereas quantum mechanically there are such fluctuations, even in a pure
state.)~\cite{barnum0,bohm2} The general idea of identifying entanglement with these quantum correlations is
widespread, e.g. see~\cite{zanardi01, zanardi02,zanardi02', zanardi04,barnum0, barnum, ortiz,peres}.

The essential idea of this experiment can be simplified a bit further: There is no need to consider the time
evolution of the prepared state; one can consider measurement immediately after the state is prepared. [15,
p.148]. A further comment is in order here, namely I have used $s_1^\zeta$ and $ s_2^\zeta$ as observables; the
fact that they are not permutation symmetric, which will probably raise objections for the electron example,
will be justified below.

Until fairly recently, emphasis has been placed on the spin state, without regard to the rest of the wave
function. But the question of how the fermionic (or bosonic) nature of particles tangles with the concept of
entanglement has been discussed more recently. Here I consider three modifications of the concept. Two of them,
due to Zanardi and coworkers~\cite{zanardi01, zanardi02, zanardi02', zanardi04}, and to Barnum et
al~\cite{barnum0, barnum, ortiz}, have in common that they apply both to spin systems and to indistinguishable
particles, plus the fact that an essential idea behind their definitions is entanglement depends on what
observables are being considered. The other, due to Schliemann \emph{et al.} and Shi~\cite{schliemann,
eckert,shi}, is specifically for fermions, and considers a state to be entangled or not, regardless of the
observables being measured. The latter definitions~\cite{schliemann, eckert,shi} have entanglement depending
only on whether the state is a single Slater determinant or not (see below).

An important point of this paper is to note that definitions of entanglement that are based on the determinantal
criterion are inconsistent with the original concept. Also we give and use in examples a definition that
embodies the essential physics of the historical concept, and is far simpler than the recent works cited.

A basic idea of the modifications~\cite{zanardi01, zanardi02, zanardi02', zanardi04,barnum0, barnum, ortiz}, as
I understand them, is that whether or not a given pure quantum state is entangled depends on what observables
are being considered as correlated or not. More precisely, \textit{a pure state $\psi$ is entangled with respect
to observables $A$ and $B$ if they are correlated in that state, i.e. if}
\begin{equation}
C_{AB}^\psi\equiv <\psi|AB|\psi>-<\psi|A|\psi><\psi|B|\psi>\ne 0;\label{1}
\end{equation}
\textit{otherwise $\psi$ is unentangled with respect to $A$ and $B$}. I'll call this definition
$\mathcal{D}$.~\footnote{This is consistent with examples in~\cite{zanardi02',zanardi04}, although there and
in~\cite{barnum0,barnum} the entanglement criterion is expressed in different,  mathematically quite
complicated, terms, which I frankly do not follow in detail. I will use the much simpler definition
$\mathcal{D}$, but will make contact with some results of~\cite{zanardi02'} (see the 2-site Hubbard model
discussion below).} Note: (i) $C_{AB}^\psi={C_{BA}^\psi}^*$, so that this definition is independent of the order
of $A$ and $B$ insofar as its being zero or not. (ii) If $\psi$ is an eigenstate of $A$, then it is unentangled
with respect to $A$ and \textit{any other} observable $B$; but $\psi$ being an eigenstate of $A$ (or $B$) is not
necessary for it to be unentangled with respect to $A$ and $B$. (iii) $A$ and $B$ need not commute. Under this
definition, \textit{any} pure state is unentangled with respect to \textit{some} observables. We will see that
this mathematically very simple definition corresponds quite directly to the essential measurements
characterizing EPR-type experiments.

 A familiar example from the spin-only class:
\begin{equation}
\chi=[\alpha_\uparrow(s_1)\alpha_\downarrow(s_2)-\alpha_\downarrow(s_1)\alpha_\uparrow(s_2)]/\sqrt{2};\label{2}
\end{equation}
$\alpha_\uparrow(s)$ and $\alpha_\downarrow(s)$ are the familiar ``spin-up" and ``spin-down" states for a
spin-1/2 particle. [To establish my notation: With $\hat{z}$ a unit vector in the up direction,
$s_i^z\alpha_\sigma(s)=\sigma\alpha_\sigma(s)$ where $\sigma$ takes on numerical values $\pm1$ corresponding to
pictorial values $\uparrow,\downarrow$; thus we take $s_i^z$ as having eigenvalues $\pm 1$, so our $s_i^z$ is
what is commonly called $\sigma_i^z$. In the usual representation, the argument $s$ takes on those eigenvalues.]
We see that this state $\chi$ is entangled with respect to $s_1^z$ and $s_2^z$ (and with respect to $s_1^\zeta$
and $s_2^\zeta$ for any direction $\zeta$), but is unentangled with respect to $\mathbf{s}^2$ and $B$, where
$\mathbf{s}=\mathbf{s}_1+\mathbf{s}_2$ is the total spin and $B$ is an arbitrary observable. The historical
definition of entanglement is the same as this definition, with the restriction that the observables be
``local", i.e. $A$ refers to one particle and $B$ to the other; $\mathbf{s}$ would be called a global
observable. The definition with respect to such global variables, while possibly of interest for some purposes,
removes the charm or ``mystery" of the historical view, which involves the idea of ``spooky action-at-a-
distance".~\cite{mermin}

Shi~\cite{shi}, Schliemann \emph{et al.}~\cite{schliemann}, and Eckert \emph{et al.}~\cite{eckert} have defined
entanglement of fermions as follows: A state that is not a single Slater determinant (i.e. not reducible to such
a determinant), is entangled; any single Slater determinant is unentangled. I will show that this definition is
not a generalization of the earlier concept, but rather is inconsistent with that concept. Note that any
normalized 2-electron Slater determinant $D$ is, by definition of the creation operators, $a^\dagger$ and $
b^\dagger$
\begin{equation}
D=a^\dagger b^\dagger|0>=\frac{1}{\sqrt{2}}\mathcal{A}a(\xi_1)b(\xi_2).\label{3}
\end{equation}
$\mathcal{A}$ is the antisymmetrizer ($\mathcal{A}f(\xi_1,\xi_2) \equiv f(\xi_1,\xi_2)-f(\xi_2,\xi_1)$), we can
take $\xi_i=\mathbf{r}_i,s_i$ in the Schr\"{o}dinger representation ($\mathbf{r}_i$ is the position of the
$i^{th}$ particle), and $a(\cdot), b(\cdot)$ are orthonormal 1-electron states.

Consider the single determinant where a single spatial function $u(\bf{r})$ (orbital) is occupied by two
electrons:
\begin{equation}
D_0 = a_\uparrow^\dagger a_\downarrow^\dagger|0>,\label{4}
\end{equation}
where $a_\sigma^\dagger$ creates a particle in the state $u(\mathbf{r})\alpha_\sigma(s)$. I consider the
Schr\"{o}dinger representation, and will not distinguish between the state vector and this representation. It is
easy to see that
\begin{equation}
D_0=u(\mathbf{r}_1)u(\mathbf{r}_2)\chi(s_1,s_2),\label{5}
\end{equation}
where $\chi$ is the singlet~(\ref{2}). In form, $D_0$ is seen to be the familiar Hartree--Fock ground state of
the He atom or the H$_2$ molecule in the molecular orbital approximation. Clearly
\begin{equation}
C_{s_1^z,s_2^z}^{D_0}=-1,\label{6}
\end{equation}
so that according to $\mathcal{D}$, the single Slater determinant $D_0$ is entangled (with respect to $s_1^z$
and $s_2^z$). Thus we've proved what we said we would, namely, the definition in
references~\cite{shi,schliemann,eckert} is inconsistent with $\mathcal{D}$, which includes the historical
definition of entanglement.

But some questions need be asked. One concerns our use of $s_1^z$ and $s_2^z$ as observables, a usage proscribed
by at least two authors: in [18, p. 388], and [15, p. 127], it is stated that any observable for a system of
identical particles must be invariant under permutations of the particles. This has some force in view of the
agreed-upon fact that there is no way in principle to distinguish between such particles in quantum mechanics.
Following a suggestion of Birge~\cite{birge}, I considered the \emph{z}-component of the spin density
\begin{equation}
s(\mathbf{R})=\sum_i s_i^z\delta(\mathbf{R}-\mathbf{r}_i),\label{7}
\end{equation}
which clearly satisfies the required symmetry. Of course~(\ref{7}) is related to the field operators
$\hat{\Psi}_\sigma(\mathbf{R})$ by $s(\mathbf{R})=\hat{\Psi}_\uparrow(\mathbf{R})^\dagger
\hat{\Psi}_\uparrow(\mathbf{R})-\hat{\Psi}_\downarrow(\mathbf{R})^\dagger \hat{\Psi}_\downarrow(\mathbf{R})$.
Taking the observables as $A=s(\mathbf{R}), B=s(\bf{R'})$ with $\mathbf{R}\ne\mathbf{R'}$, yields the same
conclusion: the correlation $C_{A,B}^{D_0}\ne 0$, so that the determinant $D_0$ is entangled with respect to the
spin density at two different spatial points. Actually, this gives a way to closely mimic the EPR experiment:
there the measurement apparatus consists of two detectors, at positions $\bf{R}$ and $\bf{R}'$. And one accepts
only coincidences where there's a particle at $\bf{R}$ and a particle at $\bf{R}'$. Thus one measures the
average $<s(\bf{R})s(\bf{R}')>$ \textit{conditional} on a particle being at $\bf{R}$ the other one being at
$\bf{R}'$. In the state $D_0$, one can see that this average is just that $<s_1^zs_2^z>$ calculated in~(\ref{6})
(which is independent of $\bf{R}$ and $\bf{R}'$). I add that the ``mystery" is present in the sense that the
distance $|\bf{R}-\bf{R}'|$ can be large enough to render any interactions negligible.~\footnote{The short range
interaction of neutrons might make them preferable for the experiment.}

The fact that working with the properly symmetric operator~(\ref{7}) yields the same result as treating the
non-symmetric $\mathbf{s}_1$ and $\mathbf{s}_2$ as observables suggests that the proscription against
considering, in general, an operator that is not permutation symmetric as an observable might be too strong. If
one interprets $\mathbf{r}_1, s_1^z$, not as the position and spin of particle \#1, but rather as the position
and spin of \textit{a} particle, it seems that this might be acceptable, and consistent with other approaches.
It would be useful if so, since, as one can see, it is formally simpler than working through the (symmetric)
spin density.

Zanardi's discussion of the 2-site Hubbard model~\cite{zanardi02'} is consistent with definition $\mathcal{D}$.
This is seen, e.g., when the interaction $U=0$. The ground state is just~(\ref{5}), with
$u(\mathbf{r})=[w_1(\mathbf{r})+w_2(\mathbf{r})]/\sqrt{2},$ the bonding orbital; $a^\dagger_\sigma$ in~(\ref{4})
is $2^{-1/2}(c^\dagger_{1\sigma}+ c^\dagger_{2\sigma})|0>$, where $c_{i\sigma}^\dagger|0>
=w_i(\mathbf{r})\alpha_\sigma$, $w_i(\mathbf{r})$ being the Wannier function at site $i$. Thus, according to
$\mathcal{D}$, $D_0$ is unentangled with respect to e.g. the $U=0$ Hamiltonian $H_{Hubb}^0$ and any other
variable (which corresponds to Zanardi's entropy $S_0=0$ at $U/4t=0$ in his FIG. 1, $\Lambda^*$), but is
\textit{entangled} with respect to the two spins $s_1^\zeta,s_2^\zeta$ (corresponding to FIG. 1, $\Lambda$).
Actually, for the purpose of the present paper, the main point is that there are variables with respect to which
\textit{this single determinant is entangled}, rather than the degree of entanglement, so that the particular
choice of such variables is beside the point. However, some preliminary results that show the possibility of
basing a definition of degree of entanglement on $\mathcal{D}$ are given in the Appendix. Incidentally, Eq. (11)
of~\cite{zanardi02'} is incorrect; it is probably just a misprint, Eq. (12) being correct for $U=0$.

I have discussed only a particular Slater determinant~(\ref{4}), which is entangled with respect to the two
spins, unentangled with respect to the positions of the particles. Instead, the determinant
$c_{1\uparrow}^\dagger c_{2\uparrow}^\dagger|0>$ is entangled with respect to the positions, unentangled wrt the
spins. More general determinants will be entangled with respect to both spins and positions. It is clear that a
Slater determinant will always be entangled with respect to to some ``local" observables (i.e.
$A(\mathbf{p}_1,\mathbf{r}_1, \mathbf{s}_1)$ and $B(\mathbf{p}_2,\mathbf{r}_2 ,\mathbf{s}_2)$), because the
antisymmetry prohibits the determinant from being a product of single-particle states. ($\mathbf{p}$ is linear
momentum.) A question then is, is it possible in any sense to talk about fermions entangled or not with respect
to such local observables? The answer is yes, under special conditions, by considering mapping of the true
fermion wave functions to non-fermion states. One example is the familiar case of a collection of hydrogen
atoms, or, simpler, Hubbard atoms, when the overlap of the Wannier functions (or the hopping parameter $t$) is
small enough (\textit{but nonzero)}; then low-lying states are governed to a good approximation by the
Heisenberg Hamiltonian, and are in $1-1$ correspondence with site-spin states, and the latter have no
permutation symmetry restrictions. Thus for some observables (e.g. low-temperature thermodynamic properties),
one can ascribe the entanglement or lack thereof in spin states to the corresponding fermion states
(see~\cite{kaplan}). An example where this correspondence is exact occurs if one is interested only in ``perfect
half-swaps" (see~\cite{kaplan2}).

In summary, the definition of fermion entanglement that includes the statement, any Slater determinant is
unentangled, has been shown to be inconsistent with the historical definition. This was done by pointing out
that experimental investigation of a particular Slater determinant~(\ref{4}) or~(\ref{5}) will lead to the EPR
type of phenomena wherein the famous peculiarly quantum correlations between the spin components of the two
``entangled" fermions occur. The discussion involved the spin density operators at two widely separated spatial
points, which formally served as observables that showed the correlation effects. It allowed a close
correspondence to the actual experiment, which measures the conditional average of the product of the spin
densities at the points of the measuring detectors, this conditional average being identical to the average of
the product of the spins of the two particles. Examples given in the text and in the Appendix show that the
question of whether a wave function is a single determinant or not is irrelevant to the question of its
entanglement. It was noted that the definition $\mathcal{D}$ of entanglement adopted here is consistent with the
idea that a true generalization of the historical concept can be made such that entanglement depends on the
observables being considered. Furthermore, $\mathcal{D}$ directly embodies the idea that entanglement means
correlation of observables, and is far simpler than other definitions which also embody this dependence on the
observables being measured.\\ \textbf{Acknowledgements}\\ I thank C. Piermarocchi, S. D. Mahanti, N. Birge and
M. Dykman for encouragement and useful discussions. Thanks to A. Mizel for important correspondence, and to R.
\vspace{.1in}Merlin for making me aware of the fermion entanglement problem.\\
\textbf{Appendix: On the Degree of Entanglement.} \\
A straightforward definition of the \textit{degree} of entanglement $E_{AB}^\psi$,
based on the previous discussion and definition $\mathcal{D}$ of whether or not a state is entangled, is simply
$|C_{AB}^\psi|$; preferable would be to normalize it by the maximum of $|C_{AB}^\phi|$ over all states $\phi$ of
interest, which is what we suggest:
\begin{equation}
E_{AB}^\psi\equiv |C_{AB}^\psi|/\max_\phi |C_{AB}^\phi|.\label{8}
\end{equation}
Then $0\le E_{AB}^\psi\le 1$, putting this quantity on the same scale as the common definition via von Neumann
entropy.~\cite{bennett}

 To get a feeling for the physical significance of this (tentative)
 definition, we apply it to the ground state
 $\Psi$ of the 2-site 2-electron Hubbard model (following Zanardi~\cite{zanardi02'}), for various choices
 of $A,B$. We put $U/(4t) = x$, where $U,t$, both $\ge0$, are the
 Coulomb repulsion and hopping parameter.
 The maximization in the denominator of ~(\ref{8}) is
 over all $\phi$ in the (6-dimensional) space
 of 2-particle states appropriate to the model. The calculation of all averages in $\Psi$ is straightforward;
 I found the calculation of the denominator in~(\ref{8}) not so straightforward.\\
(i) $\{A,B\}=\{s_1^z,s_2^z\}$ (electron spins)

 Since the ground state in this model is a product of a (symmetric) spatial state
 and the singlet, it is obvious that $<\Psi|s_1^zs_2^z|\Psi>=<\chi|s_1^zs_2^z|\chi> = -1$ for all $x$.
 Similarly $<\Psi|s_i^z|\Psi>=0$, so that
\begin{equation}
C_{s_1^z,s_2^z}^\Psi=-1, \ \mbox{independent of}\  x.\label{9}
\end{equation}
One can see~\cite{kaplan3} that this gives the maximum of $|C_{s_1^z,s_2^z}^\phi|$, so $E_{s_1^z,s_2^z}^\Psi=1$
($\Psi$ maximally entangled) for all $x$. (This includes $x=0$ where $\Psi$ is a single determinant, and, as
discussed above, is clearly appropriate to an EPR type experiment.)\\
(ii) $\{A,B\}=\{S_1^z,S_2^z\}$ (site spins)

The site spins are rather standardly defined as
\begin{equation}
S_i^z=N_{i\uparrow}-N_{i\downarrow};\label{10}
\end{equation}
$N_{i\sigma}=c_{i\sigma}^\dagger c_{i\sigma}$ are the site-spin occupation numbers. (Similar definitions are
made for the other components, of course.) One finds straightforwardly
\begin{eqnarray}
C^\Psi_{S_1^z,S_2^z}&=&-\frac{1}{1+f(x)^2},\nonumber\\
 f(x)&=&\sqrt{1+x^2}-x,\label{11}
 \end{eqnarray}
 and it can be shown~\cite{kaplan3} that $max_\phi|C^\phi_{S_1^z,S_2^z}|=1.$
(\ref{11}) is the same as~(\ref{9}) for $x\rightarrow\infty$. Interestingly, it differs for finite $x$, the
largest difference, a factor of two, occurring for $x=0$. This raises the question, under what circumstances are
the site spins observed in experiment. In the usual derivation of the Heisenberg Hamiltonian (for large $x$),
the spins that appear are the site spins. (It should be realized that this is a physics question, beyond any
question of entanglement definition.)\\
 (iii) $\{A,B\}=\{n_{\uparrow},n_{\downarrow}\}$ (bonding orbital
occupation numbers)

With
 $a_\sigma\equiv(c_{1\sigma}+c_{2\sigma})/\sqrt{2}$, the bonding orbital occupation
 numbers are defined as
$ n_\sigma=a_\sigma^\dagger a_\sigma$. One finds
 \begin{equation}
 E_{n_\uparrow,n_\downarrow}^\Psi =4g(x)[1-g(x)],\label{12}
 \end{equation}
 where
 \[g(x)=\frac{[1+f(x)]^2}{2[1+f(x)^2]}.\]
 The normalizing maximum is 1/4.~\cite{kaplan3}
 We see that $E_{n_{\uparrow},n_{\downarrow}}^\Psi = 0$ at $x=0$, as is understandable since in this
 limit, the wave function, a Slater determinant, is an eigenstate of both
 these observables. As in the discussion above of Zanardi's
 example, this noninteracting ground state is unentangled with
 respect to one set of observables, but (maximally) entangled for the other set
 ($s_1^z,s_2^z$) considered. At the other extreme,
 $E_{n_\uparrow,n_\downarrow}^\Psi\rightarrow 1$ as $x\rightarrow\infty$. Thus, according to
 the definition~(\ref{8}), the ground state is
 maximally entangled with respect to these ``non-interacting occupation
 numbers" in the strongly-interacting limit--quite reasonable.\\
 (iv) $\{A,B\} = \{N_1, N_2\}$ (site occupation numbers)

 These are defined as $N_i=N_{i\uparrow} + N_{i\downarrow}$. One
 finds easily
 \begin{equation}
 C_{N_1,N_2}^\Psi = -\frac{f(x)^2}{1+f(x)^2}.\label{13}
 \end{equation}
This $\rightarrow0$ as $x\rightarrow\infty$, so the ground state in the strongly-interacting limit is
unentangled with respect to to these site variables. Again this is because this state is an eigenstate of $N_1$
and $N_2$. Note that this state is \textit{not a single determinant, nor can it be reduced to one}. Thus we have
another example of \emph{un}entanglement of a state that is \emph{not} a single Slater determinant (the other
obvious ones are $\Psi$ for all $x > 0$, with respect to $\mathbf{s}^2$ and $B$, arbitrary $B$). In the
non-interacting limit the magnitude of~(\ref{13}) is 1/2, showing that this state, a single determinant, is
entangled with respect to these observables, but not maximally. (The normalizing maximum can be shown to be
unity~\cite{kaplan3}.)

\vspace{.2in}


\begin{thebibliography}{99}

\bibitem{bennett} C. H. Bennett, D. P. DiVincenzo, J. A. Smolin and
W. K. Wooters, \textit{Mixed-state entanglement and quantum error correction}, \emph{Phys. Rev. A} {\bf 54}
(1996), 3824.

\bibitem{zanardi01} P. Zanardi, \textit{Virtual quantum subsystems}, \emph{Phys. Rev. Lett.} {\bf 87} (2001),
077901.
\bibitem{zanardi02} P. Zanardi and X. Wang, \textit{Fermionic entanglement in itinerant systems}, \emph{J. Phys. A:Mathematical
and General;} {\bf 35} (37) (2002), 7947.
\bibitem{zanardi02'} P. Zanardi, \textit{Quantum entanglement in fermionic lattices}, \emph{Phys. Rev. A} {\bf 65} (2002),
042101.

\bibitem{zanardi04} P. Zanardi, D. A. Lidar and S. Lloyd, \textit{Quantum tensor product structures are observable induced},
\emph{Phys. Rev. Lett.} {\bf 92} (2004), 060402.
\bibitem{shi} Y. Shi, \textit{Quantum entanglement of identical particles}, \emph{Phys. Rev. A} \textbf{67} (2003),
024301.

\bibitem{schliemann} J. Schliemann, D. Loss and A. H. MacDonald, \textit{Double-occupancy errors, adiabaticity,
and entanglement of spin qubits in quantum dots.} \emph{Phys. Rev. B} {\bf 63} (2001), 085311.
\bibitem{eckert} K. Eckert, J. Schliemann, D. Bruss, and M.
Lewenstein, \textit{Quantum correlations in systems of indistinguishable particles}, \emph{Annals of Physics}
{\bf 299} (2002), 88.
\bibitem{barnum0}H. Barnum, E. Knill, G. Ortiz, , R. Somma and L.
Viola, \textit{A subsytem-independent generalization of entanglement}, \emph{Phys. Rev. Lett.} {\bf 92}, 107902
(2004).
\bibitem{barnum} H. Barnum, E. Knill, G. Ortiz and L. Viola, \textit{Generalizations of entanglement
based on coherent states and convex sets}, \emph{Phys. Rev. A} {\bf68} (2004), 032308.
\bibitem{ortiz} G. Ortiz, R. Somma, H. Barnum, E. Knill and L.
Viola, \textit{Entanglement as an observer-dependent concept: an application to quantum phase transitions},
quant-ph/0403043 v1 4 Mar 2004.
\bibitem{schrodinger} E. Schr\"{o}dinger, \textit{Die gegenw\"{a}rtige Situation in der quantenmechanik},
\emph{Naturwissenschaften} {\bf23} (1935), 823, 844.
\bibitem{einstein} A. Einstein, B. Podolsky and N. Rosen, \textit{Can quantum-mechanical description of
physical reality Be considered complete?}, \emph{Phys. Rev.}{\bf 47} (1935), 477.
\bibitem{bohm} D. Bohm, \textit{Quantum Mechanics}, Prentice-Hall (1951) sec. 22.16.
\bibitem{peres}A. Peres, \textit{Quantum Theory:Concepts and
Methods, Kluwer Academic Publishers}, Dordrecht, Boston (1993).
\bibitem{bohm2}Bohm~\cite{bohm} uses the term correlated in a
different sense for the classical case.
\bibitem{mermin} This phrase is attributed to Einstein. See \textit{The Born-Einstein Letters},
with comments by M. Born, Walker, New York (1971),
referred to in N. D. Mermin, \textit{Is the moon really there when nobody looks? Reality and the quantum
theory}, Physics Today/April, pg.38 (1985).
\bibitem{baym} G. Baym, \textit{Lectures in Physics}, W. A. Benjamin, New York (1969).
\bibitem{birge} N. Birge, private comm. (2004).

\bibitem{kaplan} T. A. Kaplan and C. Piermarocchi, \textit{Spin Swap and $\sqrt{SWAP}$ vs. Double Occupancy
in Quantum Computation}, in Proc. of INDO-US Workshop, ``Nanoscale Materials: From Science to Technology", Puri,
India, April 5-8, 2004, to appear.
\bibitem{kaplan2} T. A. Kaplan and C. Piermarocchi, \textit{Spin swap vs. double occupancy in quantum gates},
\emph{Phys. Rev. B }\textbf{70} (2004), 161311(R) .
\bibitem{kaplan3} T. A. Kaplan, unpublished work.

\end{thebibliography}
\end{document}